\documentstyle[12pt]{article}
\begin{document}
\renewcommand{\theequation}{\thesection.\arabic{equation}}

\mbox{ } \hfill  April 24th, 1994
\vspace*{2cm}
\begin{center}
{\LARGE\bf Regge Asymptotics of Scattering    }
\end{center}
\begin{center}

{\LARGE\bf with Flavour Exchange in QCD    }
\end{center}
\vspace{0.5cm}
\begin{center}
{\large\bf R. Kirschner}\\{\bf DESY - Institut f\"ur
Hochenergiephysik, Zeuthen, Germany}
\end{center}
\vspace*{2.0cm}
The contribution to the perturbative Regge asymptotics of the exchange
of two reggeized fermions with opposite helicity is investigated. The
methods of conformal symmetry known for the case of gluon exchange are
extended to this case where double-logarithmic contributions dominate
the asymptotics. The Regge trajectories at large momentum transfer are
calculated.

\newpage
\section{Introduction}

The phenomenology of processes dominated by the perturbative Regge
kinematics, the overlap of the Regge region and the range of
applicability of perturbative QCD, attracts increasing interest. The
leading singularity of the vacuum channel in perturbative QCD, the
perturbative pomeron, is well investigated \cite{BFKL} \cite{L86}.
It is expected to show up in the low $x$ asymptotics of structure
functions \cite{lowx}, in jet-inclusive final states with rapidity gaps
in deep-inelastic, hadron-hadron and other processes \cite{Mue}. More
effort has to be applied to work out the unitarity corrections to the
leading logarithmic approximation in order to calculate the true
asymptotics \cite{JB} \cite{Leff} \cite{KLS} \cite{L90} \cite{L93}
\cite{LPad}.

   The perturbative Regge singularities of the channels with quantum
number exchange, carried by quark and anti-quark of different flavour,
dominate in more specific quantities, which are more difficult to
measure in the kinematical range of interest with enough accuracy. They
should show up in the small $x$ behaviour of flavour non-singlet
structure functions, corresponding distributions with jets at rapidity
gaps and in the diffractive hadronic scattering with relatively large
momentum transfer and identified particles \cite{StB}.

The perturbative Regge asymptotics of these channels has been calculated
in the double logarithmic approximation \cite{RK81} \cite{KL83}. The
partial-wave equation in the single-logarithmic approximation has been
considered in \cite{JKw} \cite{RK86}.

The double-logarithmic contributions prevented a straightforward
extension of the methods developed for the perturbative pomeron
\cite{L86} \cite{L90} \cite{L93} \cite{Lfac}, where such contributions
do not appear.
The pomeron equation transformed into the impact parameter space shows
conformal symmetry and holomorphic factorization. These properties are
very useful in treating the multi reggeon exchange, which in the case of
only gluons gives the desired unitarity corrections to the perturbative
pomeron and the odderon \cite{GLN}.

It is important to extend the ideas developed for gluon exchange to
fermions. A treatment of multi-reggeon exchange including fermions is
needed for a complete understanding of the Regge asymptotics in
perturbative QCD. The generalization of the methods to fermions gives
a much deeper theoretical understanding. The generalization of the
effective action approach \cite{KLS} serves as a good example.

  In the present paper we show that the properties of conformal symmetry
and holomorphic factorization hold also for the two-fermion exchange.
The double-logarithmic contributions are no obstacle any more for
aplying the methods developed in \cite{L86} \cite{L90} \cite{L93}
\cite{Lfac}. Here we restrict ourselves
to the case of opposite helicities of the exchanged fermions, because
the double logarithms arise in this case only.

   The scalar $\phi^3 $ in 6 dimensions served as a toy model providing
an understanding  of how conformal symmetry may be compatible with
double logarithms \cite{RK92}. The operator representation \cite{L90}
\cite{L93} gave the appropriate framework for solving our problem. The
approach
developed in \cite{L86} \cite{KL90} has been applied for improving the
fixed coupling approximation.
The quark anti-quark Regge trajectories have been studied in the large
$N_C $ approximation in \cite{Th}.

We rederive the partial-wave equation starting from the effective
graphical rules on which the effective action is based \cite{KLS}. We
solve the equation first at vanishing momentum transfer, find the
leading Regge singularities and show, that the eigenvalue spectrum is
compatible with conformal symmetry and holomorphic factorization. Then
the operator representation in the impact parameter space is
investigated. We generalize the conformal approach to the case where the
exchanged particles correspond to conformal operators with non-vanishing
conformal weight. We derive a representation of the equation, where the
operator involved is a sum of a function of the Casimir operator of the
holomorphic conformal transformations and a function of the
corresponding anti-holomorphic Casimir operator. Finally we calculate
the series of moving Regge poles emerging from the fixed branch-points
by imposing the condition of continuity at the transition from the Regge
region to the deep inelastic region, where renormalization group
determines the behaviour.

\section{The partial-wave equation }
\setcounter{equation}{0}

  We present a short derivation of the partial-wave equation starting
from the effective graphical rules \cite{KLS}. The bremsstrahlung
contributions are already included in the effective vertex and in the
fermion trajectory. Therefore a discussion of bremsstrahlung separation
as in another derivation \cite{RK86} is not needed here.

In the logarithmic approximation the leading contribution arises from
the t-channel exchange of two reggeized fermions, which interact by
exchanging s-channel gluons. The effective vertex of gluon emission by
an exchanged fermion is given by (Fig. 1a)
\begin{equation}
T^{a}_{\alpha \alpha^{\prime} } \ \ (\phi^{* a} \kappa^{\prime *}
 + \phi^{a} \kappa^{\prime *} )
\end{equation}
 $k$ is the four-momentum and $\kappa $ its transverse (with respect to
the momenta of the incoming particles) part in complex notation ($\kappa
= \kappa_1 + i \kappa_2 $). $\phi^{a}$ and its complex conjugate stand
for the wave functions of the gluon of helicity -1 or +1, respectively.
The helicity state of the exchanged fermion is indicated by an arrow on
the line in Fig. 1;  the vertex and the propagator corresponding to the
opposite helicity are obtained by complex conjugation. The fermion
propagator (Fig. 1b) is given by
\begin{equation}
                \frac{1}{\kappa^*}
\end{equation}
and that of the s-channel gluon by  (Fig. 1b)
\begin{equation}
\langle \phi^{*} \phi \rangle  = \frac{1}{(k - k^{\prime })^2
+i\epsilon} \frac{1} {\vert (\kappa - \kappa^{\prime } \vert^2},
\ \ \ \langle \phi \phi \rangle = 0.
\end{equation}
In the fermion propagator the condition of multi-Regge kinematics that
the longitudinal part $k_+ k_-$ of the momentum squared $k^2 = k_+ k_-
-\vert \kappa \vert^2 $ is small compared to the transverse part has to
be imposed, maybe by a step function multiplying (2.2). (Compared to
\cite{KLS} in the expressions (2.1)-(2.3) we changed consistently some
factors of 2).

The trajectory of the reggeized fermion is $\frac{1}{2} + \alpha_F
(\kappa )$, where
\begin{equation}
\alpha_F (\kappa )  = - \frac{g^2 C_2 }{(2 \pi)^3}
\int \frac{d^2\kappa^{\prime }}{\vert \kappa -\kappa^{\prime }\vert^2 }
\frac{\kappa}{\kappa^{\prime }}
\end{equation}
Here a regularization has to be adopted, e.g. by extending the dimension
of the transverse space $2 \rightarrow 2 + 2\epsilon $. In this
regularization the trajectory does not depend on the helicity.

The amplitude of the flavour-exchange high-energy scattering has the
structure given by the graph Fig. 2. We shall not be concerned with the
impact factors $\Phi$ depending on the scattering particles. We
concentrate on the universal two-reggeon Green function, describing the
exchange two reggeized fermions of opposite helicity inteacting by gluon
exchange. It obeys the equation represented graphically in Fig. 3.

 We have collected the ingredients for writing the equation. The
propagator of the two reggeized fermions (the first term on r.h.s.) is
given by
\begin{equation}
\delta^{(2)} (\kappa - \overline \kappa )
[\kappa^* (q - \kappa ) (\omega - \alpha_F (\kappa ) - \alpha_F (q-
\kappa )) ]^{-1}
\end{equation}
For the amplitude Fig. 2 we use the Mellin representation with $\omega $
being the angular momentum. This simplifies in particular the
longitudinal part ($k_+^{\prime } k_-^{\prime } $) of the loop integral
in the second term. It reduces to ($\beta = 2 k_- / \sqrt s $)
\begin{eqnarray}
\int_0^{(z- \sqrt{z^2 -1})} \frac{d\beta}{\beta} \beta^{\omega } = \ \ \
\frac{1}{\omega } \ \ \ \ \ \ \ \left \lbrace \matrix {1 & \hbox{,\ \ \
\ } z \sim 1 \cr ({2 \over z} )^{\omega } & \hbox{,\ \ } z
\rightarrow \infty }  \right.
\end{eqnarray}

  $z$ is the cosine of the $t$-channel scattering angle. If $k^2 \approx
- \vert \kappa \vert^2 $ and $k^{\prime 2} \approx  -\vert
\kappa^{\prime } \vert^2 $ are of the same order then $z$ is of order 1
and the longitudinal integral yields just $\omega^{-1} $. This step is
usually done without long discussion.  However in our case the ranges of
integration where the ratio of
$\vert \kappa \vert $ and $\vert
\kappa^{\prime } \vert $ is either large or small are essential. There
$z$ is approximated by the ratio of the maximum to the minimum of
$\vert \kappa \vert $ and $\vert
\kappa^{\prime } \vert $ and the second case in (2.5) applies.

The equation Fig. 3 is written in the following form
\begin{eqnarray}
\kappa^* (q - \kappa ) [\omega - \alpha_F (\kappa ) - \alpha_F (q -
\kappa )] \ \  \tilde f(\omega ,\kappa ,\overline \kappa , q ) &=
\delta^{(2)} (\kappa - \overline \kappa )   \\   \nonumber
+\frac{g^2 C_2} {(2 \pi )^3 } \int d^2 \kappa^{\prime }
{\kappa^{\prime *} (q - \kappa^{\prime}) +
\kappa^{*} (q - \kappa)  \over \vert \kappa - \kappa^{\prime } \vert^2 }
&\left( {2 \vert \kappa \vert \vert \kappa^{\prime } \vert  \over
(\vert \kappa \vert + \vert \kappa^{\prime } )^2 } \right)^{\omega }
\tilde f(\omega , \kappa^{\prime } , \overline \kappa , q )
\end{eqnarray}
   We project on the colour singlet state in the $t$-channel. The
$SU(N)$ gauge group  generators in the vertices reduce to the factor
$C_2 = (N^2-1)/2N $ multiplying the kennel in (2.7). The second factor
in the integrand takes the effect of the upper limit in (2.5)
approximately into accout. $\omega $ can be considered as a small
parameter and therefore this factor can be approximated in different
ways. In the following form of the equation we choose an approximation
most convenient for our purposes. It is reasonable to redefine the
reggeon Green function $\tilde f $ by including the factor $\kappa^*
(q-\kappa) $. We suppress the argument $\overline \kappa $ and write an
arbitrary Born term.
\begin{eqnarray}
 \omega f(\kappa , q) = f_0 (\omega, \kappa ,q)\ \ \ \ \ \ \ \ \ \  \\
\nonumber
 + \frac{g^2 C_2 }{(2 \pi )^3 } \int \frac{d^2 \kappa^{\prime }}{\vert
\kappa -\kappa^{\prime } \vert^2 }
\left[
\left(\frac{\vert \kappa \vert }{\vert \kappa^{\prime } \vert }
\right)^{\omega }
 + \left(\frac{\vert \kappa^{\prime } \vert}{\vert \kappa \vert }
\right)^{\omega }
{\kappa^* (q - \kappa ) \over \kappa^{\prime *} (q - \kappa^{\prime }) }
\right]
f (\omega ,\kappa^{\prime } , q)
     \cr
 - \frac{g^2 C_2 }{(2 \pi )^3 } \int \frac{d^2 \kappa^{\prime }}{\vert
\kappa -\kappa^{\prime } \vert^2 }
\left( \frac{\kappa^*}{\kappa^{\prime *} } + \frac{q - \kappa }{q -
\kappa^{\prime }}  \right)
\ \ \  f(\omega , \kappa ,q)
\end{eqnarray}
The infrared divergences in the integrals at $\kappa = \kappa^{\prime }$
may be regularized by going to $ 2 + 2 \epsilon $ dimensions. Actually
the divergencies cancel in the sum of the two integrals in (2.8). The
first
integrand becomes large also in the regions $\vert \kappa^{\prime }
\vert \gg \vert \kappa \vert \gg \vert q \vert $ and $ \vert \kappa
\vert \gg \vert \kappa^{\prime } \vert \gg \vert q \vert $. However the
divergences are prevented by the $\omega $-dependent factors for ${\cal
R}e \ \omega < 0 $.

\section{Eigenvalue spectrum and conformal symmetry}
\setcounter{equation}{0}

At vanishing momentum transfer, $q = 0$, we have rotation symmetry in
$\kappa $-plane and the appropriate orthogonal basis for functions of
$\kappa $ is
\begin{equation}
\varphi_{n, \nu } (\kappa )  = \vert \kappa^2 \vert ^{-1/2 + i \nu }
\left( \frac{\kappa^{*}}{\vert \kappa \vert } \right)^n ,
\end{equation}
where $n$ runs over all integers and $\nu $ runs over the real axis. We
prefer to define the scalar product for our functions $f(\omega , \kappa
) $ with a weight , corresponding to the product of two fermion
propagators. Therefore we decompose with respect to
\begin{equation}
f_{n, \nu } = \vert \kappa \vert^{1/2} \varphi_{n, \nu } (\kappa ) .
\end{equation}
With (3.2) we diagonalize the homogeneous equation and obtain the
eigenvalue equation
\begin{equation}
\omega = \frac{ g^2 C_2 }{8 \pi^2 } \ \ \Omega (\omega , \nu , n),
\end{equation}
where
\begin{eqnarray}
& \Omega (\omega , \nu , n) = 4 \psi (1) - \cr
&  \psi (-i \nu + \frac{ \vert n \vert }{2} + \frac{\omega }{2} )
- \psi (i \nu + \frac{ \vert n \vert }{2} + \frac{\omega }{2} )
- \psi (1-i \nu - \frac{ \vert n \vert }{2} + \frac{\omega }{2} )
- \psi (1+i \nu - \frac{ \vert n \vert }{2} + \frac{\omega }{2} ),
 \cr
&\psi (z) = {d \over dz} \ln \Gamma (z) .
\end{eqnarray}
As the solution of the inhomogeneous equation (2.7) we obtain the
reggeon Green function at $q = 0$.
\begin{equation}
\tilde f(\omega , \kappa , \bar \kappa , 0) = \frac{1}{2 \pi^2}
\int_{-\infty}^{\infty }d \nu \sum_{n=-\infty }^{\infty }
{ \vert \kappa^2 \vert ^{-1 +i \nu } \left( {\kappa \over \vert
\kappa \vert } \right)^n
 \vert \bar \kappa^2 \vert ^{-1 -i \nu } \left( {\bar \kappa \over
\vert \bar \kappa \vert } \right)^{-n}  \over
\omega - \frac{g^2 C_2 }{8 \pi^2} \Omega (\omega , \nu , n) }
\end{equation}
Each term in the sum over $n$ has a square-root branch point in
$\omega$, arising from pinching the integration path by two poles
in $\nu $. The positions of the branch points are the solution of
\begin{equation}
\omega_n = \frac{g^2 C_2 }{8 \pi^2} \ \ \ \ \Omega (\omega_n, 0, n).
\end{equation}
The dominant ones are at $n = 0 $,
\begin{equation}
\omega_0 = \left( \frac{g^2 C_2 }{2 \pi^2} \right)^{1/2} ( 1 +
{\cal O}(g^2) ),
\end{equation}
and at $n = \pm 1$,
\begin{equation}
\omega_1 = \frac{g^2 C_2}{\pi^2} (\ln 2 -\frac{1}{2} )  \ \ \
(1 + {\cal O}(g^2) ).
\end{equation}
All other singularities are less important, $\omega_n < 0 , \vert n
\vert > 1$. (3.7) reproduces the result of the double logarithmic
approximation \cite{KL83}. The double logarithmic contributions show up
essentially at $n = 0 $ and lead only to small corrections at $ n \not=
0$.

In the case of exchanging two reggeized gluons (perturbative pomeron)
the properties of conformal symmetry and holomorphic factorization have
been observed \cite{L86} \cite{Lfac}.For these properties to hold it
is
neceesary that the spectrum $\Omega (\omega , \nu ,n) $ can be written
as a sum of two terms, one being a function of the eigenvalues $m(1-m)$
of the holomorphic and the other of the eigenvalues $\tilde m (1- \tilde
m )$ of the antiholomorphic conformal Casimir operator, where
\begin{equation}
m =\frac{1}{2} + i \nu +\frac{n}{2},\ \ \
\tilde m =\frac{1}{2} + i \nu - \frac{n}{2}.
\end{equation}
$n$ enters (3.4) as $\vert n \vert$. A property of $\psi (z)$ allows to
rewrite the eigenvalue function as a combination of $\psi$-functions
avoiding absolute values.
\begin{eqnarray}
& \Omega (\omega , \nu , n) = 4 \psi(1) - \cr      & \frac{1}{2} \{
\psi (m + \Delta )
+ \psi (1- m + \Delta )
+ \psi (m - \Delta )
+ \psi (1- m - \Delta )    \cr
& + \psi (\tilde m + \Delta )
+ \psi (1- \tilde m + \Delta )
+ \psi (\tilde m - \Delta )
+ \psi (1- \tilde m - \Delta )  \},
\end{eqnarray}
where $\Delta = (1- \omega )/2 $. Further we have
\begin{equation}
2 \psi (1) - \psi (m +\Delta ) - \psi (1 - m - \Delta ) =
\chi_{\Delta} ( m(1-m) ) ,
\end{equation}
with the notation
\begin{equation}
\chi_{\Delta} ( x ) = \sum_{\ell = 0}^{\infty }
\left( \frac{2(\ell + \Delta ) + 1 }
{(\ell + \Delta ) (\ell + \Delta +1 ) + x }
- \frac{2}{\ell + 1} \right).
\end{equation}
This allows to write the eigenvalue function as
\begin{eqnarray}
&\Omega (\omega , \nu , n) =    \cr  &\frac{1}{2} \left(
\chi_{\Delta} ( m(1-m) )   +
\chi_{- \Delta} ( m(1-m) ) +
\chi_{\Delta} (\tilde m (1-\tilde m ) ) +
\chi_{- \Delta} (\tilde  m (1-\tilde m ) \right).
\end{eqnarray}

We have shown that the spectrum has the desired properties. This means,
that the equation (2.8) can be represented up to non-leading terms in a
conformally symmetric form allowing holomorphic factorization.

\section{Representation in impact parameter space}
\setcounter{equation}{0}

Instead of finding an integral representation in  impact parameters as
in \cite{L86} we shall work with operators as in \cite{L90}. The
momenta
in (2.7) translate into derivatives with respect to the coordinates
$x_1, x_2$
\begin{equation}
\kappa^* \rightarrow \partial_1, \ \ \ \ \ q - \kappa \rightarrow
\partial^*_2,
\end{equation}
the transverse momentom factor in the gluon propagator
translates as
\begin{equation}
 \frac{1}{\vert \kappa - \kappa^{\prime \vert^2 }}
\rightarrow \ \ \  - \ln \vert x_{12}^2 \vert
\end{equation}
where $x_{12} = x_1 - x_2$,
and the fermion trajectory
\begin{equation}
\alpha ( \kappa )
\rightarrow - \ln \vert \partial^2 \vert \ \  + 2 \psi (1)
\end{equation}
(4.2) and (4.3) can be understood in dimensional regularization, where
the pole term $\sim \epsilon^{-1}$ is omitted in both expressions.
Indeed the pole terms cancel in the operator below, because there are no
infrared divergences in (2.7).

We write the equation (2.7) as
\begin{equation}
\omega \  \tilde f = \ \ \tilde f_0 \ + \  \frac{g^2 C_2 }{8 \pi^2}
\ \ {\cal  H}_{F\overline F}  \ \tilde f
\end{equation}
In the impact parameter representation the partial wave depends on four
points in the impact parameter space, $\tilde f = \tilde f (\omega ,
x_1, x_2; \bar x_1, \bar x_2) $. The operator acts on $x_1, x_2 $ only.
It can be read off directla from (2.7) using (4.1) - (4.3).
\begin{eqnarray}
{\cal  H}^{(\omega )}_{F\overline F}        &= - (\partial_1
\partial^*_2 )^{-1} \left(
(\partial_1 \partial^*_2 )^{\omega /2} \ln \vert x_{12}^2 \vert \ \
(\partial_1 \partial^*_2 )^{1-\omega/2 }   +
(\partial^*_1 \partial_2 )^{- \omega /2} \partial_1 \partial^*_2
\ln \vert x_{12}^2 \vert \ \ (\partial^*_1 \partial_2 )^{\omega/2 }
\right)                 \nonumber  \\
& - \ln \vert \partial^2_1 \vert
- \ln \vert \partial^2_2 \vert  + 4 \psi (1)
\end{eqnarray}
The effect of the upper limit of the integration is included in a way
similar to (2.8). ${\cal  H}_{F\overline F} $ allows holomorphic
factorization, because it decomposes into the sum $ H_F + \tilde H_F $,
where $H_F $ acts only on the holomorphic coordinates $x_1 ,x_2 $ and
$\tilde H_F $ only on $x_1^* ,x_2^* $.
\begin{eqnarray}
H_F^{(\omega )} =
& - (\partial_1 )^{-1 + \omega /2 } \ln x_{12} (\partial_1 )^{1 - \omega
/2 } - \ln \partial_1   \cr
& - (\partial_2 )^{- \omega /2 } \ln x_{12} (\partial_2 )^{ \omega
/2 } - \ln \partial_2 + 2 \psi (1)
\end{eqnarray}
$\tilde H^{(\omega )}_F $ is obtained from (4.6) by complex conjugation
and interchanging the indices 1 and 2.

We present several forms for $H_F $ in the limiting case $\omega = 0 $.
Mathematically this limit makes sense for the operator although in the
corresponding equation one would loose the dominant double-logarithmic
contributions. The following discussion is analogous to the one for the
two-gluon exchange \cite{L93}.

 Applying the commutation relation to the first term in (4.6) (at
$\omega = 0 $) we obtain
\begin{equation}
 H_F^{(0)} = -2 \ln x_{12} - \ln \partial_1 - \ln \partial_2
-\partial_1^{-1} (x_{12} )^{-1}  + 2 \psi (1) ,
\end{equation}
which can be written by an analogous application of the commutation
relation as
\begin{equation}
 H_F^{(0)} = -2 \ln x_{12} - \ln \partial_2 - x_{12} \ \ \
\ln \partial_1  \ \ \ \
 x_{12}^{-1}  + 2 \psi (1) .
\end{equation}
Using the operator relation
\begin{equation}
 \ln (x^2 \partial ) - \ln x = \partial^{-1} \ \  \ln x \ \  \partial
- \ln \partial
\end{equation}
one obtains from (4.6)
\begin{equation}
H_F^{(0)} = -\ln (x_{12}^2 \partial_1) - \ln \partial_2 + 2 \psi (1)
\end{equation}
The last form is convenient for studying the behaviour under conformal
transformations. In particular under inversions ${\cal I}$ \
$H_F^{(0)} $ transforms in the following way.
\begin{equation}
{\cal I} \ \  H_F^{(0)} \ \  {\cal I} = x_2 \ H_f^{(0)} \ x_2^{-1}
\end{equation}

The forms (4.6) and (4.8) imply that the transposed operator can be
obtained from $H_F^{(0)} $ by two different similarity transformations,
\begin{equation}
H_F^{(0) T} = \partial_1 \ H_F^{(0)} \ \partial_1^{-1} =
P_{12} x_{12}^{-1} \  H_F^{(0)} \ \   x_{12} P_{12} .
\end{equation}
Here $P_{12} $ denotes the operator permuting $x_1 $ and $x_2 $. From
(4.12) we understand, that the following operator commutes with
$H_F^{(0)}$.
\begin{equation}
A_F = P_{12} x_{12} \partial_1 , \ \ \ \ [A_F, H_F^{(0)} ]  = 0.
\end{equation}

The operator relation (4.9) can be derived by observing that
\begin{eqnarray}
x^2 \partial &=& \Gamma (x \partial ) \ x \ (\Gamma (x \partial ))^{-1},
\cr
 \partial &=& ( \Gamma (x \partial + 1) )^{-1} \ x^{-1} \ \Gamma (x
\partial + 1) .
\end{eqnarray}
The operators in the similarity transformation are actually determined
only up to periodic functions with the period 1 of the same argument.
This ambiguity matters, if we apply (4.14) to logarithms.

(4.14) implies that both sides of (4.9) are equal to $\psi (x \partial
)$. It  also allows to write (4-10) as a sum of $\psi $-functions.
\begin{equation}
H_F^{(0)} = - \frac{1}{2} \left(
\psi (D_1 +1 ) + \psi ( -D_1 )   +
\psi (D_2 +1 ) + \psi ( -D_2 )   \right)
+ 2 \psi (1)
\end{equation}
We have used the notation $D_1 =x_{12} \partial_1, D_2 = x_{12}
\partial_2 $.
There is an uncertainty in the last form of the operator due to the
periodic function ambiguity in (4.14).

The operator $H_F^{\omega }$ at $\omega \not= 0 $ is more complicated.
As
it stands, it does not have the symmetry properties (4.11) and (4.13).
On the other hand we know, that up to non-leading terms it approximates
an operator with nice symmetry properties.

Applying (4.14) and going the analogous steps which led to (4.15) we
obtain, that up to non-leading contributions
\begin{eqnarray}
H_F^{(0)} &=
- \frac{1}{2} \partial_1^{-\omega /2} \left( \psi (D_1 +1) + \psi (-D_1
) \right) \partial_1^{\omega /2}
- \frac{1}{2} \partial_2^{-\omega /2} \left( \psi (D_2 +1) + \psi (-D_2
) \right) \partial_2^{\omega /2}  +2 \psi (1)   \cr
 &= - \frac{1}{2} \left(
\psi (1 + D_1 - \frac{\omega }{2}) + \psi (1 + D_2 - \frac{\omega }{2})
\right) \cr &- \frac{1}{2} \left(
\psi (- D_1 + \frac{\omega }{2}) + \psi (- D_2 + \frac{\omega }{2})
\right) + 2 \psi (1).
\end{eqnarray}

\section{Using conformal symmetry}
\setcounter{equation}{0}

Now we consider the diagonalization of ther equation (4.4) by means of
conformal symmetry. We have seen that the equation with $H_F^{(0)} $ is
symmetric and we know that it can be made symmetric by non-leading
corrections also with $H_F^{(\omega )}$. The functions $\tilde f$ can be
expanded in conformal 3-point functions,
\begin{eqnarray}
E (x_{10}, x_{20} ) =
\langle \phi^{(\Delta_1 , \tilde \Delta_1 )}  (x_1)
 \phi^{(\Delta_1 , \tilde \Delta_1 )}(x_2 )
{\cal O}^{(m, \tilde m)} (x_0) \rangle   \cr
= \left( \frac{x_{12}}{x_{10} x_{20}} \right)^{m-\Delta_1 -\Delta_2}
x_{10}^{-2 \Delta_1 } x_{20}^{-2 \Delta_2 }
\cdot \left( \frac{x^*_{12}}{x^*_{10} x^*_{20}} \right)^{\tilde m
- \tilde \Delta_1 - \tilde \Delta_2} x_{10}^{* -2 \tilde \Delta_1 }
x_{20}^{* -2 \tilde \Delta_2 } , \cr
\Delta_{1/2} = \frac{1}{2} (\delta_{1/2} + s_{1/2}),      \ \ \
\tilde \Delta_{1/2} = \frac{1}{2} (\delta_{1/2} - s_{1/2}),    \cr
m = \frac{1}{2} ( \delta + n), \ \ \ \ \tilde m = \frac{1}{2} (\delta -
n ).
\end{eqnarray}
$\delta_{1/2}, \delta $ are the scaling dimensions and $s_{1/2}, n $ are
the conformal spins of the operators in the points $ x_{1/2 } ,x_0 $.
The
behaviour of $ H_F^{(0)} $ under inversion (4.11) shows that $ \tilde f
$ transforms non-trivially. We arrive at the conclusion that the
operators in $x_{1/2} $ corresponding to the exchanged fermions carry
conformal weights
\begin{equation}
\Delta_1 = 0,  \ \ \Delta_2 = \frac{1}{2},\ \ \
\tilde \Delta_1 = \frac{1}{2}, \ \ \tilde \Delta_2 = 0.
\end{equation}
The 6 generators of the conformal transformations in the corresponding
representation on functions of $x_1,  x_2 $  have the form
\begin{eqnarray}
M_{12}^0 &= x_1 \partial_1 + x_2 \partial_2 + \Delta_1 + \Delta_2, \ \ \
M_{12}^- = \partial_1 + \partial_2, \cr
M_{12}^+ &= x_1^2 \partial_1 + 2 \Delta_1 x_1 + x_2^2 \partial_2 + 2
\Delta_2  x_2,
\end{eqnarray}
with obvious modification for the anti-holomorphic generators $\tilde
M_{12}^{\pm}, \tilde M_{12}^0 $.
They obey the commutation relations
\begin{equation}
[M_{12}^0, M_{12}^{\pm} ] = \pm M_{12}^{\pm } , \ \ \ \ \
[M_{12}^+, M_{12}^- ] = -2 M_{12}^0 .
\end{equation}
We need the explicite form of the Casimir  operator
\begin{equation}
C_{\Delta_1 ,\Delta_2 } = - ( M_{12}^0 )^2 \ \ + \ \
\frac{1}{2} (M_{12}^+ M_{12}^- + M_{12}^- M_{12}^+ )
\end{equation}
for our case (5.2)
\begin{equation}
C_{0, \frac{1}{2} } = \ x_{12}^2 \partial_1 \partial_2
-  x_{12} \partial_1 + \frac{1}{4} .
\end{equation}
We notice a remarkable property, which may be related to supersymmetry,
\begin{equation}
A_F^2 = - C_{0,\frac{1}{2} } + \frac{1}{4}.
\end{equation}
Therefore the commutativity (4.13) of $H_F^{(0)} $ with $A_F $ is
just another way to express its conformal symmetry.

The properties of orthogonality and completeness of the conformal
3-point functions discussed in \cite{L86} \cite{KL90} for $ \Delta_1 =
\Delta_2 =  0 $ hold also in the general case.  We shall not write
these
relations and the expansion of the solution of (2.7), (4.4), the reggeon
Green function, in conformal 3-point functions here, because the form is
like in \cite{L86} \cite{KL90}.  We have the twofold overcomplete
orthogonal set (5.1) of functions of $x_1, x_2 $ parametrized by $\delta
= \frac{1}{2} + i \nu , -\infty < \nu < \infty, $ by $n$ running over
all integers and by the position $x_0 $ running over the whole complex
plane. This becomes evident, if we transform to the $\rho q $
representation, where $\rho = x_{12} $ and the momentum transfer $q$ is
the Fourier conjugate to $R = (x_{10} + x_{20} )/2$.  In the
asymptotics
$ \rho q \ll 1 $ we come close to the rotation symmetric situation (3.1)
and obtain for the conformal 3-point functions the simple approximation
\begin{eqnarray}
& \tilde E_{\delta_1, \delta_2 ,s_1 , s_2 }^{n, \nu }  =
\vert \rho \vert ^{1- \delta_1 - \delta_2 }
\left({\rho \over \rho^*} \right)^{\frac{s_1 + s_2}{2}}
r_{n, \nu }  \cdot                \cr
& \left[
\vert \rho \vert ^{- 2 i \nu }
\left({\rho \over \rho^*} \right)^{\frac{n}{2}}
+ e^{i \delta (n, \nu )}
\vert q^2 \rho \vert ^{ 2 i \nu }
\left({q^{* 2} \rho \over q^2 \rho^*} \right)^{- \frac{n}{2}}
\right]
\ \ \left( 1 + {\cal O}(\rho q) \right) .
\end{eqnarray}
The constants $r_{n, \nu } $ and $\delta (n, \nu )$ are known. In the
following we need the phase
\begin{eqnarray}
  e^{- \delta (n, \nu ) } =
  2^{4 i \nu }  \ \ \ \ \ \ \ \ \ \ \ \ \ \ \ \ \ \  \;\;\;\;\;\;\;\;
\;\;\; \nonumber \\
 { (2 i \nu + \vert n \vert ) \Gamma^2(2 i \nu + \vert n
\vert )
\Gamma (- i \nu + \frac{\vert n \vert + 1}{2} - \frac{|s_1 - s_2 | }{2})
\Gamma ( i \nu + \frac{\vert n \vert + 1}{2} + \frac{|s_1 - s_2 |}{2} )
\over {\rm c.c. }         }
\end{eqnarray}

The conformsymmetric operator $H_{F {\it sym}}^{(\omega )} $, a
non-symmetric approximation to which is given by (4.16), is a function
of the Casimir operator $C_{0, \frac{1}{2}} $ (5.6). This holds in
particular in the asymptotics $\rho q \ll 1 $, where the approximation
(4.16) is good. On the other hand in this asymptotics
\begin{equation}
D_1 + D_2  \sim \rho q  \ll 1
\end{equation}
 We start from (4.16), use the formula
\begin{equation}
 \psi (1) - \psi (z) = \sum_{\ell = 0 }^{\infty } \left (
\frac{1}{\ell + z } - \frac{1}{\ell +1 } \right )
\end{equation}
and (5.6)
\begin{equation}
C_{0, \frac{1}{2} } =  \  D_2 D_1 + \ \frac{1}{4} .
\end{equation}
In the approximation (5.10) we have
\begin{eqnarray}
\frac{1}{\ell + 1 +D_1 - \frac{\omega }{2} } +
\frac{1}{\ell + 1 +D_2 - \frac{\omega }{2} }    = \cr
\frac{2 \ell + 2 -\omega }{(\ell + 1 - \frac{\omega }{2})^2 + D_2 D_1 }
=          \cr
\frac{2 (\ell + \Delta ) + 1 }{
(\ell + \Delta ) (\ell + 1 + \Delta ) + C_{0, \frac{1}{2}} }
\end{eqnarray}
In this way we obtain the expression of $H_{F {\it sym }}^{(\omega )} $
in terms of the conformal Casimir operator.
\begin{equation}
H_{F {\it sym }}^{(\omega )}  = \frac{1}{2} \left(
\chi_{\Delta } (C_{0, \frac{1}{2} } )   +
\chi_{- \Delta } (C_{0, \frac{1}{2} } )  \right)
\end{equation}
where the notation (3.12) has been used with $\Delta = (1 -\omega )/2 $.
Together with the analogous expression for $\tilde H_{F {\it
sym}}^{(\omega )} $ this confirms the result (3.13) about the
eigenvalues spectrum.

\section{Regge trajectories at relatively large momentum transfer}
\setcounter{equation}{0}

The expression of the reggeon Green function in terms of conformal
3-point functions obtained from (4.4) generalizes (3.5) to arbitrary
momentum transfer $t = q^2 $. It has a form analogous to (3.5) and the
singularities in $\omega $ arise just in the same way. The perturbative
Regge singularities do not move with $t$ in the leading logarithmic
approximation with a fixed coupling $g$.

The approximation of a fixed coupling is reasonable in the channels
without double logarithms, if $t$ and $\kappa^2 $ are of the same
order.
Then this scale determines the coupling, because essentially all momenta
sqared are of this order.  In channels dominated by double-logarithmic
contributions the region of applicability of the fixed coupling
approximation is more restricted. The transverse momentum integrals may
get essential contributions from $\kappa^{\prime 2} \gg \kappa^2 \sim t
$. However for $\omega $ not small, $\omega \gg g^2 $, such
contributions are suppressed. Under this condition on $\omega $ also in
this case the fixed coupling approximation can be justified.

Actually the eqation has been derived in the approximation of small
$\omega $. At small coupling there is an intermediate region, where
$\omega $ is small enough for this approximation and still provides
enough suppression of large transverse momenta. Indeed, as we have seen,
the leading singularity in the double logarithmic channel (5.7) is at
$\omega_{0} \sim g \gg g^2 $.

In both cases the fixed coupling approximation is not true anymore, if
$\kappa $ becomes large (or $\rho $ small ), which is the deep-
inelastic region. There the dependence on $\rho $ ( or on $\kappa $ ) is
determined by the renormalization group equation. The equation (2.7),
(4.4) has to be modified to become compatible with the renormalization
group.

We write (4.4) in the $\rho q $ representation approximated for $\rho q
\ll 1 $.
\begin{eqnarray}
\omega f(\rho) = \frac{g^2 C_2}{8 \pi^2 } \ \ \ \Omega (\omega , \hat
\nu , n)   \ \ \ f(\rho) , \cr
\hat \nu = i {\partial \over
\partial \ln (\vert \rho \vert^2 \Lambda^2 ) }
\end{eqnarray}
We restrict ourselves to the channels with conformal
spin $ n = 0, {\pm 1} $,
where the leading singularities (3.7), (3.8) arise. The solution have
the form (5.8) with the conformal weights (5.2). Inverting $\Omega $ as
a function of $\nu $ (6.1) can be written as
\begin{equation}
\vert \rho \vert^2 \frac{\partial}{\partial \vert \rho \vert^2 } \ \
f(\rho ) = \nu (\frac{g^2}{4 \pi }, \omega , n) \ \ f(\rho )
\end{equation}
The renormalization group leads to replacing
\begin{equation}
\frac{g^2}{4 \pi } \rightarrow \alpha_S (1/ \rho ) =
\frac{-b}{\ln (\vert \rho \vert^2 \Lambda^2 )}
\end{equation}
in the scaling dimension $ \nu $ in (6.2). $b$ is the leading
coefficient in the Gell-Mann - Low function.

Instead of studying (6.2) with the replacement (6.3) we solve (6.1) with
the replacement (6.3). The asymptotics of the solutions at $\rho
\rightarrow 0 $ is the
same in both cases and is given by the saddle point approximation of the
integral
\begin{eqnarray}
f(\rho) =
\int_{- \infty }^{\infty } d\nu
\exp \left[ -i \nu \ln (\vert \rho \vert^2 \Lambda^2 )   +
\frac{i b C_2}{2 \pi \omega }
\int_0^{\nu } \Omega (\omega , \nu^{\prime }, n) d\nu^{\prime }
\right]
\end{eqnarray}
The saddle point equation coincides with (3.3) after the replacement
(6.3).

In the region where $\rho q \ll 1 $ but still
\begin{equation}
\alpha_S (q^2) \ln (\vert \rho q \vert^{-2} ) \ll 1
\end{equation}
the solutions of the original equation (6.1) and the one modified by
(6.3) should match. In particular the condition for the phases to match
(modulo $\pi $ ) restricts the possible values of $\nu $ to a discrete
set $\nu^{(r)}(\omega ) , \ \ \ r$ integer.
\begin{eqnarray}
\nu^{(r)} \ln (\vert \rho \vert^2 q^2 )
+ \frac{1}{2} \delta (n, \nu^{(r)} ) + \pi r  =   \cr
 -{\pi \over 4} + \nu^{(r)} \ln (\vert \rho \vert^2 \Lambda^2 )
+ \frac{b C_2 }{2 \pi \omega }
\int_0^{\nu^{(r) }} \Omega (\omega , \nu^{\prime } ,n) d\nu^{\prime }
\end{eqnarray}
In the solution obeying the condition of continuity between the
perturbative Regge (fixed coupling) region and the deep inelastic region
instead of the integral over $\nu $ like in (3.5) there is a sum over
$r$ with the solutions of (6.6) $\nu^{(r)} $ inserted. The
singularities in $\omega $ are now a series of poles at $\omega_n^{(r)}
$, the solution of
\begin{equation}
\omega_n^{(r)} = \alpha_S (q^2) \ \ \frac{C_2}{2 \pi } \ \ \
\Omega (\omega_n^{(r)} , \nu^{(r)}_n , n).
\end{equation}
We solve (6.6), (6.7) for $n = 0, \pm 1 $ within our approximation. For
this we use the approximation of $\Omega (\omega , \nu , n) $ (3.4) for
small $\omega , \nu $ disregarding terms beyond the second order in
$\omega , \nu $ and also the value of the phase $\delta (n, \nu ) $ at
$\nu = 0$.

At $n = 0$ we obtain
\begin{equation}
\omega^{(r)}_0 =
\left( \frac{2 C_2 \alpha_S (q^2)}{\pi } \right)^{1/2}
\left[ 1 - \left( {\alpha_S (q^2) \over C_2 } \right)^{1/3} \pi
 \left(\frac{3}{8 b} (r +\frac{1}{4} ) \right)^{2/3}
\right]
\end{equation}
and at $n = \pm 1$
\begin{eqnarray}
\omega_1^{(r)} = \frac{2 \alpha_S (q^2 ) C_2 }{  \pi }  (2 \ln 2 - 1 )
\left[ 1 - (\alpha_S (q^2) )^{2/3}
\left( {21 \ \zeta (3) - 4 \over 2 \ln 2 - 1 } \right)^{1/3}
\left(\frac{ 3\pi}{2 b} (r + \frac{1}{4} ) \right)^{2/3}
\right]
\end{eqnarray}

Applying this method the Regge trajectories were  calculated for the
channel $n = 0$ earlier in \cite{Th}. The phase $\delta (n, \nu )$ was
left undetermined there.

Only a discrete set of scaling dimensions $ \nu^{(r)} $ allows the
continuation of the  solution from  the perturbative Regge region  to
the deep inelatic region. Instead of the branch point we obtain an
infinite
series of Regge poles moving with $t = q^2$. In the region of relatively
large $\vert t \vert $,  where the perturbative leading logarithmic
calculation gives a reasonable estimate, the $t $
dependence comes through the renormalized coupling.

\section{Discussion}

We have extended the methods of conformal symmetry in the Regge
asymptotics to the flavour exchange channels of QCD. Here the leading
contributions to the perturbative Regge asymptotics arise from the
exchange of two reggeized fermions ( quark and anti-quark of different
flavour) and it is determined by double logarithmic contrbutions, if the
fermions carry opposite helicities.

We have shown that the eigenvalue spectrum of the partial-wave equation
obeys the necessary conditions for conformal symmetry and factorization.
Therefore the equation can be written in a conformsymmetric form. We
have obtained this symmetric form using the operator representation in
impact parameter space. The equation is written in terms of an operator,
which is a sum of a function of the holomorphic plus a function of the
anti-holomorphic Casimir operators of the linear conformal
tranformations.

We have obtained the leading perturbative Regge singularities in the
flavour exchange channels. The singularity in the channel with conformal
spin $n=0$ arises from double-logarithmic contributions and it dominates
the channels with higher conformal spin. The result for the position of
this branch point in the complex angular momentum plane coincides with
the one obtained in the double-logarithmic approximation. This means,
that in the leading $\ln s $ approximation  there are no essential
corrections to the double-logarithmic result. As a practical consequence
our equation ( $\ln \frac{1}{x} $ approximation ) and the LAP equation
\cite{LAP}
($ \ln Q^2 $ approximation ) predict the same low $x$ asymtotics for the
non-singlet structure functions.

We are able to calculate the singularities also in channels of other
conformal spins $n$. They are determined essentially  by
single-logarithmic contributions. The case of $n = \pm 1 $ is of
interest because the branch point is located at positive angular
momentum. This contribution is a correction to the one of $n = 0 $.
There may be situations where the contribution of $n  = 0$  is
suppressed
due to the impact factors, i.e. the corresponding Regge singularity may
have  a small coupling to the scattering particles.

We have discussed the conditions of validity of the fixed coupling
calculations and the necessary modifications when going towards the deep
inelastic region. As a result the fixed branch points are converted into
series of moving Regge poles. We have calculated the Regge trajectories
in the channels $n = 0 $ and $ n = \pm 1 $ at relatively large momentum
transfer. Earlier we observed the appearance of Regge poles by including
the running coupling into the simple non-linear partial wave equation
obtained in the double logarithmic approximation \cite{RK83}. It remains
to be analyzed, how the two approaches ( for $n = 0$ dominated by double
logarithms) compare in detail.

The results about the trajectories give no information on the slope, the
intercept of the meson trajectories or on the meson masses. However it
becomes clear, that the Regge singularity structure in the meson
channels as well as in the vacuum channel is more complicated in QCD
than assumed usually in phenomenology. Any model of non-perturbative
effects resulting in a definite modification of the equation in the
region of small transverse momenta will produce by an analogous
procedure  predictions on the intercept and on the slope.

The operator representation in impact parameter space, which was
used first in \cite{L90} \cite{L93} for the gluon exchange, has been
worked out for
the application to the flavour exchange channels.  The conformal method
has been extended to the case where the exchanged reggeons are
attributed non-vanishing conformal spin and to the treatment of
double-logarithmic contributions. Now all two-reggeon interaction
operators, which appear in the partial-wave equation in the general case
of multiple exchange of reggeized quarks and gluons, can be treated in
the same way \cite{RK94}. The results about the integrability of the
equation for an arbitrary number of exchanged gluons in the large $N$
approximation \cite{LPad}   generalize to the exchange of gluons and
fermions.

\newpage

$\mbox{ }$ \\
{\Large\bf Acknowledgements} \\
$\mbox{ }$ \\
The author is grateful to L.N. Lipatov and L. Szymanowski for
discussions.

\newpage
\noindent{\Large\bf Figure captions}
\vspace{2cm}

\begin{tabular}{ll}

Fig. 1 & Effective graphical rules. a) effective vertex of gluon  \\
       &  emission, b) propagator of the exchanged fermion,     \\
       &  c) propagator of the s-channel gluon.                 \\
       & \\
Fig. 2 & Amplitude high-energy scattering. The blobs denoted by \\
       & $\Phi $ stand for the impact factors and the blob $f$ stands \\
       & for the reggeon Green function. \\
       &  $p_A $ and $p_B $ are the momenta of incomimg particles. \\
       & \\
Fig. 3 & Equation for the two-reggeon Green function. \\
\end{tabular}

\newpage


\vspace*{-2cm}
\setlength{\textheight}{24cm}
\setlength{\evensidemargin}{0.0mm}
\setlength{\oddsidemargin}{0.0mm}
\footheight0.0pt
\textwidth20cm
\input FEYNMAN

\begin{picture}(40000,21000)
\drawline\fermion[\E\REG](3000,10000)[6000]
\drawarrow[\W\ATBASE](4500,10000)
\drawline\photon[\N\REG](6000,10000)[3]
\put(4000,8000){$\kappa $}

\put(8000,8000){$\kappa^{\prime } $}

\drawline\fermion[\E\REG](14000, 10000)[12000]
\drawarrow[\W\ATBASE](18000,10000)
\put(20000,8000){$ \kappa $}

\drawline\photon[\N\REG](35000, 8000)[4]
\put(36000, 10000){$ k - k^{\prime } $}
\put(6000,4000){\sl a}

\put(20000,4000){\sl b}
\put(35000,4000){\sl c}
\put(40000,2000){\sl Fig.1 }

\end{picture}

\vspace*{1cm}

\begin{picture}(40000,21000)
\drawline\scalar[\N\REG](5000, 0)[3]
\drawline\scalar[\N\REG](5000, 16000)[3]
\put(5000,11000){\oval(3000,10000)}
\put(4500,10500){$ \Phi_A $}
\drawline\fermion[\E\REG](6500,8500)[7000]
\drawline\fermion[\E\REG](6500,13500)[7000]
\put(9500,7000 ){$ k $}
\drawline\fermion[\E\REG](13500,6000)[5000]
\drawline\fermion[\N\REG](\pbackx,\pbacky)[10000]
\drawline\fermion[\W\REG](\pbackx,\pbacky)[5000]
\drawline\fermion[\S\REG](\pbackx,\pbacky)[10000]
\put(15500,10500){$ f $}
\drawline\fermion[\E\REG](18500,13500)[7000]
\drawline\fermion[\E\REG](18500,8500)[7000]
\put(21500,7000){$ \overline k $}
\drawline\scalar[\N\REG](27000, 0)[3]
\drawline\scalar[\N\REG](27000, 16000)[3]
\put(26000,22500){$ p_B + q $ }
\put(27000,11000){\oval(3000,10000)}
\put(26500,10500){$ \Phi_B $}
\put(4500,-1000){$ p_A $}
\put(26500, -1000){$ p_B $}
\put(40000,-1000){\sl Fig. 2 }

\end{picture}

\vspace*{1cm}

\begin{picture}(50000,21000)
\drawline\fermion[\E\REG](3000,5000)[5000]
\drawline\fermion[\N\REG](\particlebackx,\particlebacky)[6500]
\drawline\fermion[\W\REG](\particlebackx,\particlebacky)[5000]
\drawline\fermion[\S\REG](\particlebackx,\particlebacky)[6500]
\put(5000,8000){$ f $}
\drawline\fermion[\W\REG](3000,6250)[3000]
\drawline\fermion[\W\REG](3000,10250)[3000]
\drawline\fermion[\E\REG](8000,6250)[3000]
\drawline\fermion[\E\REG](8000,10250)[3000]
\put(13000,8250){$ = $}
\drawline\fermion[\E\REG](16000,6250)[5000]
\drawline\fermion[\E\REG](16000,10250)[5000]
\put(23000,8250){$ + $}
\drawline\fermion[\E\REG](32000,5000)[5000]
\drawline\fermion[\N\REG](\particlebackx,\particlebacky)[6500]
\drawline\fermion[\W\REG](\particlebackx,\particlebacky)[5000]
\drawline\fermion[\S\REG](\particlebackx,\particlebacky)[6500]
\put(34000,8000){$ f $}
\drawline\fermion[\W\REG](32000,6250)[5000]
\drawline\fermion[\W\REG](32000,10250)[5000]
\drawline\fermion[\E\REG](37000,6250)[3000]
\drawline\fermion[\E\REG](37000,10250)[3000]
\drawline\photon[\N\REG](29000,6250)[4]
\put(40000,0){\sl Fig. 3}

\end{picture}

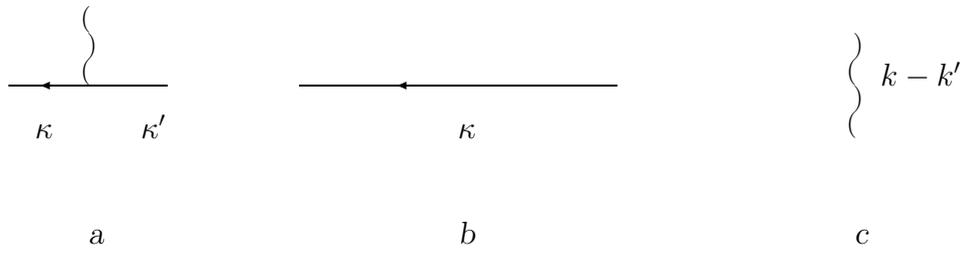
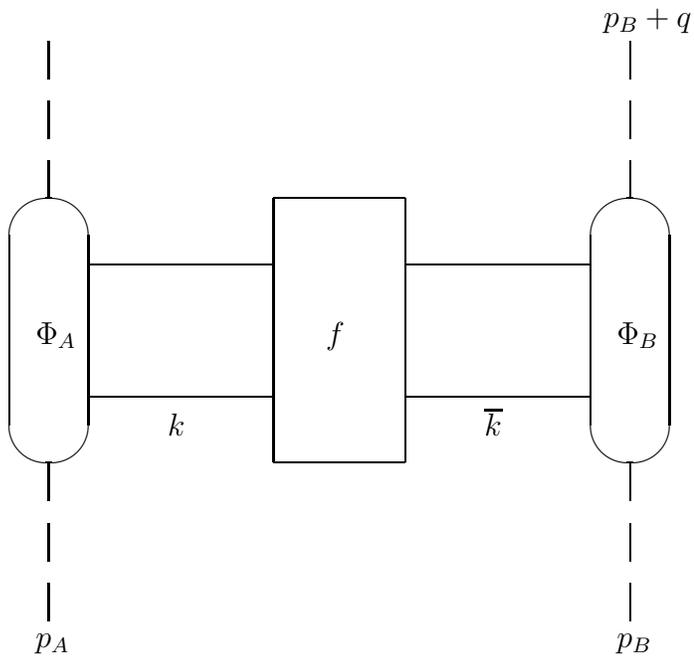
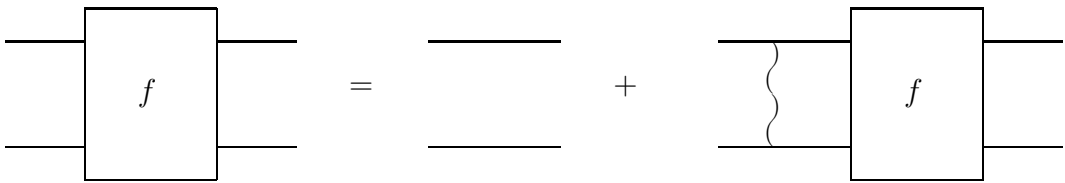
\end{document}